\documentclass[epj,referee]{svjour}
\usepackage{latexsym}
\usepackage{graphics}
\begin{document}
\title{Bethe-Heitler cross-section for
very high photon energies and
large muon scattering angles 
}
%
%
\author{Gerhard Baur\inst{1} \and Albrecht Leuschner\inst{2}
}                     
%
%
\institute{
  Forschungszentrum J\"ulich, Institut f\"ur Kernphysik,
  D-52425 J\"ulich \\ \email{G.Baur@fz-juelich.de} \and
  Deutsches Elektronen-Synchrotron DESY, Notkestrasse 85, 
  D-22607 Hamburg \\ \email{Albrecht.Leuschner@desy.de}
}
\date{Received: date / Revised version: date}
%
\abstract{
The cross-section for the process $\gamma + A \rightarrow \mu^+ + \mu^- + X$
is studied where the photon energy is of the order of several hundreds
of GeV and where one of the leptons is scattered to large angles. 
This is of practical importance for muon shielding calculations at 
future linear electron colliders.
In addition to the photon pole contribution which was previously considered
especially by Y.S.Tsai, we identify another component due to the muon pole
(equivalent photon and equivalent muon approximation). This is
discussed following the usual Bethe-Heitler formula. Then we give
practically useful formulae for inclusive lepton (muon) production
along with some numerical examples.
\PACS{
      {13.60.-r}{Photon and charged lepton interactions with hadrons}   \and
      {13.60.Hb}{Total and inclusive cross-sections 
                 (including deep inelastic processes)}   \and
      {25.20.Lj}{Photoproduction reactions}
     } 
} 
\authorrunning{G.Baur, A.Leuschner}
\titlerunning{Bethe-Heitler cross-section for very high photon energies and
large muon scattering angles}
\maketitle
\section{Introduction}
\label{Introduction}
\begin{figure}
\resizebox{0.50\textwidth}{!}{
  \includegraphics{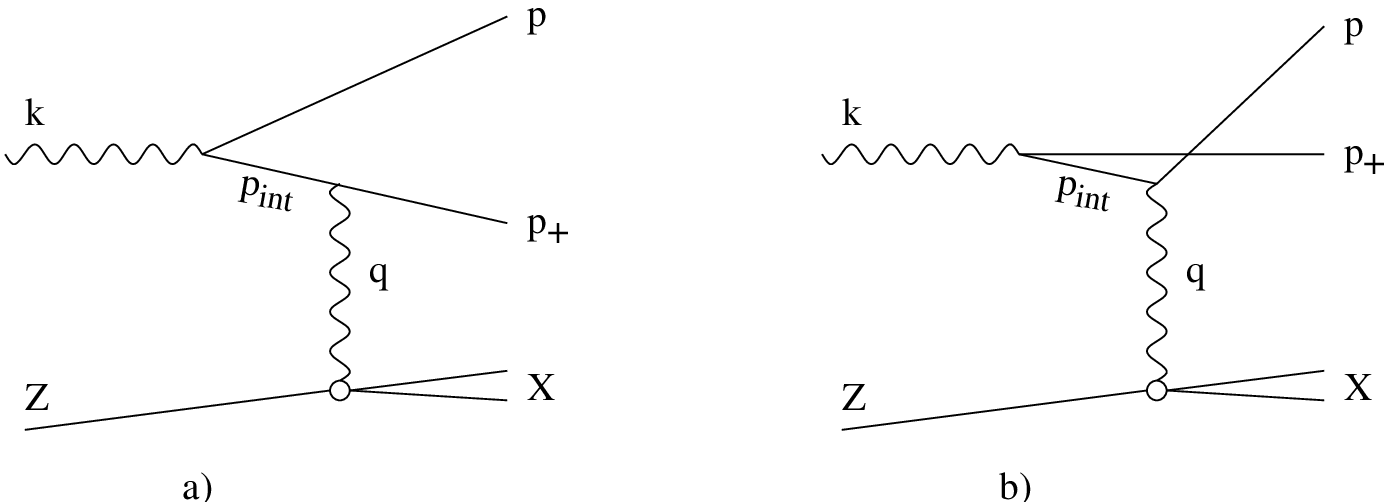}
}
    \caption{Feynman graphs describing the process 
             \mbox{$\gamma+A \rightarrow p + p_+ + X$}}
    \label{fig:intro1}
\end{figure}

This paper arose out of a practical question. At future Linear
Colliders like TESLA \cite{CDR97} electron and positron beams
of several hundreds of GeV and high beam powers of some 10~MW
have to be absorbed after having passed the interaction region.
In this process, high energy photons are produced which in turn
give rise to high energy muons. In order to estimate the muon radiation
dose at earth surface above the Linear Collider \cite{BAUR98} pair production
cross-sections for large angles are necessary. This is a well known
problem which has been studied many times, most elaborately probably
by Y.S.Tsai \cite{TSAI74}. An exact lowest order formula is given there,
which corresponds to the graphs shown in figs.\ref{fig:intro1}
a) and b).
The information about nuclear structure is fully contained in the 
electromagnetic structure functions $W_1$ and $W_2$. However,
this formula is very complicated and hard to evaluate practically,
especially if one has to integrate over the unobserved lepton with
fourmomentum $p_+$
(e or $\mu$, we are here concerned only with muons). Therefore
practical calculations were done using the Weizs\"acker-Williams
(or equivalent photon) approximation. This corresponds to the kinematical
situation where the square of the momentum of the exchanged photon,
$q^2 = -Q^2$, is specially small ($q^2 \approx 0$). In many situations,
this is the dominant distribution. In our studies we found another
kinematical situation to be important. It corresponds to the muon pole,
where the intermediate muon is close to its mass-shell (``equivalent
muon approximation''). This has been studied before \cite{CHEN75}
(see also \cite{BFK73}) and we adopt the formulations of these authors.
\paragraph{}
Starting from the usual formula for the Bethe-Heitler process
(where infinitely heavy point-like nuclei are assumed) we explain
in chapter \ref{CrossSection} the general features of the pair
production process and its important limiting cases. Then we discuss
the general case, where structure effects are taken into account (along
with the effects due to the finite mass, or recoil effects). The 
information needed as an input is sufficiently well known, specially
from electron scattering. In chapter \ref{NumericalExamples} we provide
some illustrative examples, together with discussions. Our conclusions
are given in chapter \ref{conclu}.
%
%
\section{Cross-section for $\gamma + A \rightarrow \mu^+ + \mu^- + X$}
\label{CrossSection}
\subsection{Bethe-Heitler cross-section revisited. 
            Photon and muon pole contributions}
\label{Bethe}
The cross-section for the Bethe-Heitler process is calculated in many
textbooks corresponding to the graphs of fig.\ref{fig:intro1}.
Assuming that the nucleus is infinitely heavy and point-like
one obtains the following formula for the differential cross-section
(we use the natural units $\hbar=c=1$) 
\cite{LALI86}
\begin{eqnarray}
  \label{equ:cross1}
  d\sigma &=& \frac{8 \alpha^3 Z^2 m^2 }{\pi k^3 Q^4}
            \; E_+ \, E dE  \\
         &  & \left\{ 
            -\frac{\delta_+^2}{(1+\delta_+^2)^2}
            -\frac{\delta^2}{(1+\delta^2)^2}
            +\frac{k^2}{2E_+E} 
                  \frac{\delta_+^2+\delta^2}{(1+\delta_+^2)(1+\delta^2)}
            \right. \nonumber \\
         & & \;\;  + \left. \left( \frac{E_+}{E} 
                    +\frac{E}{E_+} \right)
             \frac{\delta_+ \delta \cos(\phi)}{(1+\delta_+^2)(1+\delta^2)}
            \right\} \, \delta_+ d\delta_+ \, \delta  d\delta \, d\phi 
            \nonumber
\end{eqnarray}
where m denotes the muon mass, $k=E_+ + E$ 
is the photon energy, 
$E_+,E$ the energies of the outgoing muons and
$\phi$ is the angle between the planes spanned by the photon and
the outgoing muons. 
We have $\delta_+=\theta_+ E_+ /m$, $\delta=\theta E /m $, where
$\theta_+, \theta $ are the scattering angles of the outgoing muons, and
the momentum transfer is given by
\begin{equation}
  \label{equ:cross2}
  Q^2/m^2 = \delta_+^2 + \delta^2 + 2 \delta_+ \delta \cos(\phi)
            + m^2 \left( 
             \frac{1+\delta_+^2}{2E_+}
           + \frac{1+\delta^2}{2E}
             \right)^2
\end{equation}
Typical scattering angles are given by $\theta \cong m/ E$,
i.e. $\delta \cong 1$.
Now we are interested in the case where one of the muons (say $\mu^-$) 
scatters to large angles
and we integrate over the angles of the
other one $(\theta_+, \phi)$, i.e. we have
\begin{equation}
  \label{equ:cross3}
  \delta \gg 1
\end{equation}
In this integration, a large contribution will come from the region
where  $Q^2$ is as small as possible. This is the case for
\begin{equation}
  \label{equ:cross4}
  \delta_+ \cong \delta \qquad \mbox{and} \qquad \phi \cong \pi
\end{equation}
i.e. $\mu^+$ and $\mu^-$ scatter to opposite sides with about equal
transverse momentum. In this case eq.(\ref{equ:cross2}) leads to
\begin{equation}
  \label{equ:cross5}
  Q_{min}^2 \cong m^4 (1+\delta^2)^2 \left( 
        \frac{k}{2 E_+ E} \right)^2
\end{equation}
and one uses the Weizs\"acker-Williams approximation as it is elaborated
in \cite{TSAI74}.
\paragraph{}
In addition, there is another region of integration which can become
important,
\begin{equation}
  \label{equ:cross6}
  \delta \gg 1 \quad , \quad  0 \le \delta_+ \le \delta_{+,max} 
       \quad \mbox{and} \quad 0 \le \phi < 2\pi
\end{equation}
where the choice of $\delta_{+,max}$ is discussed below.
In this case the momentum transfer is given by
\begin{equation}
  \label{equ:cross7}
  Q_{mp}^2 \cong m^2 \delta^2 = (\theta E)^2
\end{equation}
which is generally much larger than $Q_{min}^2$. The paranthesis
in equ.(\ref{equ:cross1}) is simplified and one obtains for
the differential cross-section (integrated over the angles of the
$\mu^+$, $u \equiv \delta_+^2$, $u_{max} = \delta_{+,max}^2$)
\begin{eqnarray}
  \label{equ:cross8}
  \frac{d^2 \sigma}{d\Omega dE} &=& \frac{4 Z^2}{\pi} \; 
        \frac{\alpha^3 E_+ E^3}{k^3 Q_{mp}^4} 
        \nonumber \\
  & &   \int\limits_0^{u_{max}} du \left( 
        \frac{-u}{(1+u)^2}
        + \frac{k^2}{2E_+ E (1+u)} 
        \right)
\end{eqnarray}
The integral in equ.(\ref{equ:cross8}) diverges logarithmically.
Following \cite{CHEN75} (see also \cite{BFK73}) we put 
$\delta_{+,max} = \theta E/m \gg 1$ and we find
\begin{equation}
  \label{equ:cross9}
  \frac{d^2 \sigma}{d\Omega dE} \cong \left[ 
             \frac{4 Z^2 \alpha^2}{\theta^4 E^2} \right] \,
             \frac{\alpha}{\pi} \, \ln \left( \theta \, \frac{E}{m} \right)\,
                      \frac{E^2+(k-E)^2}{k^3}                    
\end{equation}
where $E_+=k-E$ was used.   This is in agreement 
with \cite{BFK73} (see especially equ.(13)) and with \cite{CHEN75}
(see eqs.(6)-(8)). The derivation of equ.(\ref{equ:cross9}) actually
depends on the assumption that $\delta_+ \ll \delta$.
In the next subsection
we give a simple physical meaning to equ.(\ref{equ:cross9}) in a more
general context.
\subsection{Effects of finite nuclear size, nuclear and nucleon structure}
For large scattering angles, $\delta \gg 1$, the photon and muon pole
contributions are separated in phase space, 
since $\delta_+ \cong \delta \gg 1$ for the equivalent photon approximation
(EPA) and $\delta_+ \le 1$
for the equivalent muon approximation (EMA). 
Therefore we can add these two terms incoherently
to obtain the full inclusive cross-section
\begin{equation}
  \label{equ:cross10}
  \frac{d^2 \sigma}{d\Omega dE}
                     = \left(\frac{d^2 \sigma}{d\Omega dE}\right)_{EPA}
                     + \left(\frac{d^2 \sigma}{d\Omega dE}\right)_{EMA}
\end{equation}
The relative importance of the two terms depends on the special kinematical
values and the $Q^2$-dependence of the corresponding structure functions.
\paragraph{}
The photon pole contribution was extensivly discussed in \cite{TSAI74}
and we follow this procedure and do not have to go into details here.
There are coherent nuclear scattering, described by the elastic form-factor
of the nucleus and, for larger values of $Q_{min}^2$ (i.e. larger
scattering angles), incoherent contributions due to the scattering
from individual nucleons\footnote{There is also a ``quasi-elastic''
contribution where the Pauli-suppression effect on the knocked out
nucleon is included.}. In the equivalent photon approximation, the 
effects of nucleon structure are entirely contained in the function $\chi$,
see eqn.(2.19) of \cite{KIM73}
\begin{eqnarray}
  \label{equ:cross11}
   \chi &=& \frac{1}{2M_i}\, 
                   \int\limits_{M_i^2}^{\infty} dM_f^2  
                   \int\limits_{Q_{min}^{\prime 2}}^{Q_{max}^2} 
                   \frac{dQ^2}{Q^4}
                   \left[ \,(Q^2-Q_{min}^{\prime 2}) \cdot 
                         W_2(Q^2,M_f^2) \right.  \nonumber \\
  & & \hspace{2.5cm} \left. + 2 Q_{min}^{2} \cdot W_1(Q^2,M_f^2) \,\right]   
\end{eqnarray}
The mass of the hadron system in the initial state is denoted by $M_i$, the
mass of the produced final state by $M_f$, and 
\begin{equation}
  \label{equ:cross12}
   Q_{min}^{\prime 2} = Q_{min}^2 + 2 \Delta \sqrt{Q_{min}^2}
   \quad \mbox{where} \quad \Delta=\frac{M_f^2 - M_i^2}{2 M_i} .
\end{equation}
The expression for $Q_{min}^2$ (see our eq.(\ref{equ:cross5})) 
is in accord with 
the corresponding expression given in App.A of \cite{KIM73}. 
The electromagnetic structure functions are
denoted by $W_2$ and $W_1$. The elastic contribution is well described
by the usual dipole parametrization (see \cite{TSAI74}). In this reference,
a parametrization of the inelastic contribution $M_f \not= M_i$
is also given (``meson production form-factor''). We suggest a somewhat
different approach, which takes the resonant character of the structure
function for $Q^2 \approx 0$ into account. For small enough $Q^2$
(the important region of the integration in eq.(\ref{equ:cross11})) 
the structure functions are directly related to the cross-section for real
(transverse) photons $\sigma_{\gamma p}(k)$ (the scalar part vanishes
as $Q^2 \rightarrow 0$). One has
\begin{eqnarray}
   \label{equ:cross13}
     W_2(Q^2,M_f^2) &\cong& \frac{Q^2}{4 \pi^2 \alpha}\;
                          \frac{2 M_i}{M_f^2 - M_i^2} \;
                          \sigma_{\gamma p}(M_f) \; f(Q^2)  \\
     W_1(Q^2,M_f^2) &\cong& \frac{1}{4 \pi^2 \alpha}\;
                          \frac{M_f^2 - M_i^2}{2 M_i} \;
                          \sigma_{\gamma p}(M_f) \; f(Q^2)  \nonumber
\end{eqnarray}
where a form-factor $f(Q^2)$ is introduced. The cross-section 
$\sigma_{\gamma p}$ for real photons is dominated by nucleon resonances,
most prominently by the $\Delta$-resonance. We take the 
$Q^2$-dependence of $f(Q^2)$ to be the same as for the elastic proton
form-factor, i.e. we chose a dipole form
\begin{equation}
  \label{equ:cross14}
  f(Q^2)= \frac{1}{(1+Q^2/\Lambda^2)^4}
  \quad \mbox{where} \quad \Lambda^2 = 0.71 \, \mbox{GeV}^2 \, .
\end{equation}
Note, however, that a somewhat stronger fall-off with $Q^2$ was found
experimentally \cite{BAR68}.
\paragraph{}
By applying the equivalent photon approximation to the full Bethe-Heitler
expression, one neglects the muon pole contribution. In many cases this
is justified because the $Q^2$-value involved in the muon pole contribution
is usual much higher compared to the one occuring in the equivalent photon 
approximation. Form-factor effects emphasizing the low $Q^2$-values tend
to make the muon pole contribution small. On the other hand, in the
deep-inelastic scattering region scaling sets in and the structure
functions do not decrease any more with increasing $Q^2$.
\paragraph{}
The muon pole contribution is described in \cite{CHEN75},\cite{BFK73}
(equivalent muon approximation). After integration over the unobserved
$\mu^+$ (which is scattered to small angles), the scattering process
factorizes into an ``equivalent muon spectrum'' and the scattering
cross-section of the muon on the target. This muon can be considered
as a ``parton'' inside the photon. It is moving in the direction
of the photon with an energy fraction $x=E^{\prime}/k$ of the photon,
where $E^{\prime}$ corresponds to the muon energy in the intermediate state.
The inclusive Bethe-Heitler cross-section is now obtained as 
(see eq.(25) of \cite{CHEN75})
\begin{equation}
  \label{equ:cross16}
  \left( \frac{d^2 \sigma}{d\Omega dE} \right)_{EMA} =
                         \int\limits_{x_o}^1 dx \;
                         F_{\gamma}^{\mu}(k,x) \;
        \frac{d^2 \sigma(\mu N \rightarrow \mu^{\prime} X)}{d\Omega dE}(kx)
\end{equation}
This equivalent muon spectrum
is given by \cite{CHEN75}
\begin{equation}
  \label{equ:cross15}
  F_{\gamma}^{\mu}(k,x) = \frac{\alpha}{\pi}\;
                               \ln \left( \frac{k}{m} \right)
                               \left[ x^2 + (1-x)^2 \right]
\end{equation}
A more refined expression is obtained by replacing
\begin{equation}
  \label{equ:cross15a}
   \ln \left( \frac{k}{m} \right) \qquad \mbox{by} \qquad
   \ln \left( \theta \, \frac{x k}{m} \right)  \quad \mbox{with} \quad
     \theta \gg \frac{m}{x k}                       
\end{equation}
(see also \cite{CHEN75} and \cite{BFK73}).
The muon-nucleon inclusive scattering cross-section is given by
\begin{equation}
  \label{equ:cross17}
    \frac{d^2 \sigma(\mu N \rightarrow \mu^{\prime} X)}{d\Omega dE} =
    \left( \frac{d \sigma}{d\Omega} \right)_{Mott}
      \cdot \left( W_2 + 2 W_1 \tan^2(\theta/2) \right)
\end{equation}
where the Mott cross-section is given by
\begin{equation}
  \label{equ:cross18}
      \left( \frac{d \sigma}{d\Omega} \right)_{Mott} =
             \frac{\alpha^2}{4(k x)^2} \;
             \frac{\cos^2(\theta/2)}{\sin^4(\theta/2)}
\end{equation}
where the muon energy is given by $E^{\prime}=xk$. The kinematical limit $x_o$
in eq.(\ref{equ:cross15}) is given by \cite{CHEN75}
\begin{equation}
  \label{equ:cross19}
  x_o = \frac{M_N E}{M_N k - 2 k E \sin^2(\theta/2)}
\end{equation}
in the present notation. 
Equ.(\ref{equ:cross16}) is rewritten as
\begin{eqnarray}
  \label{equ:cross20}
  \left( \frac{d^2 \sigma}{d\Omega dE} \right)_{EMA}&=& 
                             \frac{\alpha^2}{4k^2 \sin^4(\theta/2)} 
        \cdot \\   & & \hspace{-1.3cm}
         [ W_2^{\gamma}(k,E,\theta) \cos^2(\theta/2) 
  + 2W_1^{\gamma}(k,E,\theta) \sin^2(\theta/2) ] \nonumber 
\end{eqnarray}
with
\begin{equation}
  \label{equ:cross21}
  W_{1,2}^{\gamma}(k,E,\theta) = \frac{\alpha}{\pi}\;
                               \int\limits_{x_o}^1 dx \;
                               \ln\left(\theta \, \frac{x k}{m}\right)
                               \frac{x^2 + (1-x)^2}{x^2} \; W_{1,2}(\nu,Q^2)
\end{equation}
where $\nu=xk-E$ and $Q^2=4x k E \sin^2(\theta/2)$ are
the usual variables used in deep inelastic scattering. In order
to evaluate the integral in eq.(\ref{equ:cross21}) we have to know
the structure function on the ray
\begin{equation}
  \label{equ:cross22}
  Q^2 = 4(\nu+E)E \sin^2(\theta/2)
\end{equation}
For an infinitely heavy and point-like target proton we have
$W_2 = \delta(xk-E)$. The integration 
eqs.(\ref{equ:cross20}),(\ref{equ:cross21})
leads to the same cross-section as it was discussed
in section \ref{Bethe}, eq.(\ref{equ:cross9}) (for $Z=1$), if one
applies the small angle approximation ($\theta \ll 1$).  
Now one sees that eq.(\ref{equ:cross9})
factorizes into the Mott scattering cross-section
(with $E=E^{\prime}=xk$) and the equivalent muon number 
of eq.(\ref{equ:cross15}).
\paragraph{}
In principle, these structure functions are well known from deep
inelasic lepton scattering \cite{KEN91},\cite{PAR98}. For a
rough estimate of the order of magnitude of the effect we use
a simplified approach: for the rather low values of $Q^2 \ge 1 \mbox{GeV}^2$
and $M_f \ge 2.6 \mbox{GeV}$ an approximate scaling behaviour sets in
\cite{KEN91}. We have 
\begin{equation}
  \label{equ:cross23}
   \nu W_2(\nu,Q^2) = F_2(x_q)  
\end{equation}
where $x_q=Q^2/(2 \nu M_N)$ and $F_2$ is independent of $Q^2$.
We also have
\begin{equation}
  \label{equ:cross24}
  2 M_N \, W_1(\nu,Q^2) = F_1(x_q)
\end{equation}
and the Callan-Gross relation
\begin{equation}
  \label{equ:cross25}
  F_2(x_q)= 2 x_q \, F_1(x_q) .
\end{equation}
%
%
\section{Numerical Results and Discussion}
\label{NumericalExamples}
\begin{figure*}
\resizebox{1.05\textwidth}{!}{%
  \includegraphics{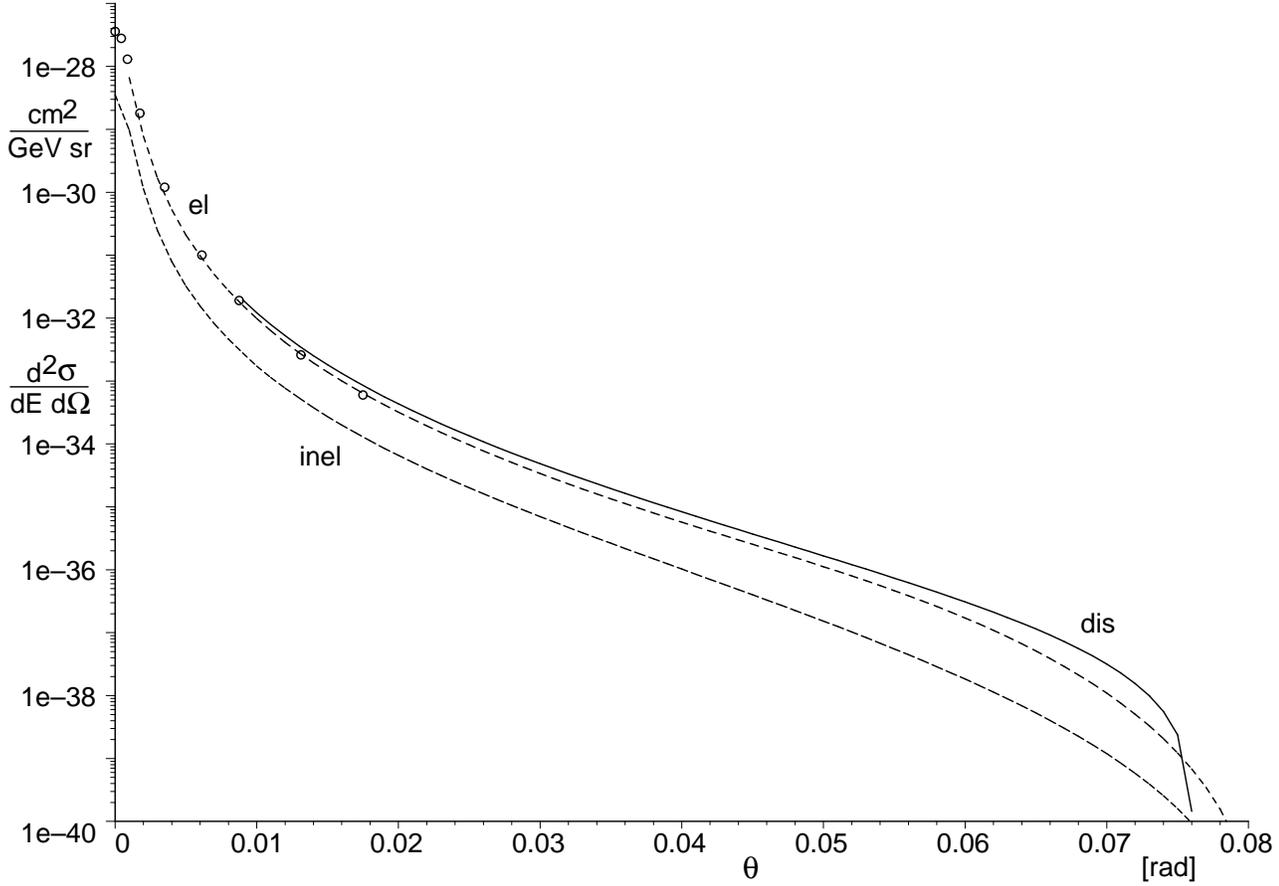}
}
\vspace{-1.0cm}
\caption{Bethe Heitler cross-sections on a proton at a photon energy of 200~GeV and
a muon energy of 120~GeV as a function of the polar angle~$\theta$.
Solid line - deep inelastic (dis), dashed lines - elastic (el) 
and inelastic (inel). For further explanations see text.}
\label{fig:res1}       
\end{figure*}

For small scattering angles,the Bethe-Heitler cross-section
is dominated by coherent nuclear scattering. With increasing angles, 
the effect 
of the form-factor of the nucleus will set in and the cross-section 
will decrease rather rapidly. The nuclear form-factor is characterized by a 
rather soft scale $\Lambda^2_{nucleus}=\hbar c/R = 0.005 \, \mbox{GeV}^2$
for R=3~fm. The corresponding scale for a nucleon is 
$\Lambda^2_N=0.71\,\mbox{GeV}^2$. For large angles, 
the incoherent scattering from the nucleons will
take over. Apart from Pauli-blocking effects,
the nucleus is just an assembly of 
Z protons and N neutrons at rest (we can neglect Fermi motion).
This is all very well and extensively described in \cite{TSAI74} 
and we can concentrate on the incoherent contributions from the nucleons.
\paragraph{}
 The elastic nucleon contribution is well described within 
the equivalent photon approximation \cite{TSAI74}. 
We take the usual dipole form-factors.
We have also checked the muon pole contribution for the elastic case.
Since the momentum transfer $Q^2_{min}$ (see eq.(\ref{equ:cross5})) 
is much smaller
than $Q^2_{mp}$ (see eq.(\ref{equ:cross7})) the strong decrease 
of the dipole form-factor with $Q^2$ renders the
muon pole contribution negligible in this case.
\begin{table}[bh]
\caption{Numerically calculated integrals of eqs.(\ref{equ:res1}).}
\label{tab:num1}      
\begin{center}
\begin{tabular}{|r|r|r|r|r|}
\hline\noalign{\smallskip}
$M_1$~[GeV] & $M_2$~[GeV] & $M_f$~[GeV] & 
 $\tilde{W_2}$~[$\mbox{GeV}^{-2}$] &  $\quad \tilde{W_1}$     \\
\noalign{\smallskip}\hline\noalign{\smallskip}
 1.11  & 1.35 & 1.24 & 2.66  & 0.30  \\
 1.35  & 1.62 & 1.49 & 1.24  & 0.62 \\
 1.62  & 2.05 & 1.82 & 1.00  & 1.67  \\
 2.05  & 2.55 & 2.31 & 0.65  & 3.56 \\
 2.55  & 3.25 & 2.81 & 0.62 & 10.09 \\
\noalign{\smallskip}\hline
\end{tabular}
\end{center}
\end{table}
\paragraph{}
Inelastic contributions for small $Q^2$
are dominated by nucleon resonances. We use eqs.(\ref{equ:cross11}) 
and (\ref{equ:cross13}) to calculate the inelastic contribution.
We take the cross-section $\sigma_{\gamma p}$ for real photons from
experimental data \cite{PAR98}. This cross-section is dominated by
nucleon resonances (mainly the $\Delta$) in the GeV region, followed
by a structureless continuum. For the integration over $M$ 
the cross-section $\sigma_{\gamma p}$ is regarded as a 
sequence of 5 resonances at the center of mass energies 
$M_f$ as listed in tab.\ref{tab:num1}. The first 2 are the real
resonances. The relative widths of all the regions are kept approximately the
same. The following integrals corresponding to equ.(\ref{equ:cross13}) were
numerically calculated for each region:
\begin{eqnarray}
   \label{equ:res1}
     \tilde{W_2} &=&  \frac{1}{4 \pi^2 \alpha}\;
                          \int \limits_{M_1}^{M_2} 2 M \, dM\;
                          \frac{1}{M^2 - M_i^2} \;
                          \sigma_{\gamma p}(M)   \\
     \tilde{W_1} &=&  \frac{1}{4 \pi^2 \alpha}\;
                          \frac{1}{(2 M_i)^2} \;
                          \int \limits_{M_1}^{M_2} 2 M \, dM\;
                           (M^2 - M_i^2) \;
                          \sigma_{\gamma p}(M)   \nonumber \\
     M_f         &=&     \int \limits_{M_1}^{M_2} dM\;
                            M \, \sigma_{\gamma p}(M) \left/  
                         \, \int \limits_{M_1}^{M_2} dM\;
                                   \sigma_{\gamma p}(M) \right. \nonumber
\end{eqnarray}
They are listed in tab.\ref{tab:num1}.
A parametrization of the contribution of the $\Delta$-resonance to
$W_2$ is given by Chanfray et al. \cite{CHAN93}, our results are in
qualitative agreement.
At large angles the resulting 
cross-section  behaves
very similarly to the elastic cross-section 
(shown in fig.\ref{fig:res1}, dashed lines)
as expected from the $Q^2$-dependencies of the form factors. 
It should be noted here that proton and neutron behave similarly
as it is indicated by comparing the $\gamma p$ with
the $\gamma d$ cross-section \cite{PAR98}.
%
%
\paragraph{}
We concentrate now on the muon pole contribution.
Due to the quark-parton structure of the
nucleon, the form factor does not decrease any more for high $Q^2$ values.
This is the reason why the muon pole contribution is important.
For (roughly) $Q^2 > 1\,\mbox{GeV}^2$ and $M_f > 2.6\,\mbox{GeV}$
scaling sets in \cite{KEN91}. Following \cite{OHN94} we put
\begin{equation}
  \label{equ:res2}
   F_2(x_q)=\rho\,\ln(1/x_q) \qquad \mbox{with} \qquad \rho=0.16   
\end{equation}
This is a very rough approximation which has the merit 
that it leads to an analytical expression. 
For our present purposes this seems sufficient. 
Using eqs.(\ref{equ:cross20})-(\ref{equ:cross25}) and 
(\ref{equ:res2}) we get for the
deep inelastic (dis) cross-section
\begin{eqnarray}
  \label{equ:res3}
  \left( \frac{d^2 \sigma}{d\Omega dE} \right)_{dis}&=& 
                             \frac{\alpha^3}{4 \pi k^2 \sin^4(\theta/2)} 
                             \cdot \rho \cdot
                             \int\limits_{x_o}^1 dx \;
                             \ln\left(\theta \, \frac{x k}{m}\right) \cdot \\  
    & & \hspace{-2.0cm}  \cdot    \frac{x^2 + (1-x)^2}{x^2} \cdot
                 \ln \frac{2M_N (x k - E)}{x k E \theta^2} \cdot
                 \left( \frac{1}{x k - E} + \frac{x k - E }{x k E} \right)
 \nonumber 
\end{eqnarray}
\paragraph{}
At the expense of using more computer time,
more sophisticated expressions for the structure function
$F_2(x_q)$ can be inserted. 
The structure functions of the neutron are also known
\cite{FRI91},\cite{AUB87}. A simple approximation is given in \cite{FEY72}
$F_{2,neutron}=(1-x_q)F_{2,proton}$, valid for $x_q < 0.75$. 
For $x_q > 0.75$ is $F_{2,neutron} \cong F_{2,proton}$
(see fig.8 of \cite{FRI91}).
Since $F_{2,proton}$ is always larger than $ F_{2,neutron}$,
one would only overestimate the cross section, 
when treating all nucleons as protons.
In the following we restrict ourselves to the Bethe-Heitler 
process on a proton. As a typical case we take $k=200\,\mbox{GeV}$.
The energy of the outgoing muon is, following Tsai \cite{TSAI74}, 
taken to be 120~GeV.
For smaller angles,we can also compare our calculations to Tsai.
We give 3 contributions
\begin{description}
\item[The elastic production.] 
Here the photon pole contribution dominates over 
the muon pole contribution. For the latter the momentum transfer 
is considerably larger and form-factor effects render it negligible
(this was checked, but we do not need to show it here). 
Fig.\ref{fig:res1} showes our elastic cross-section in good 
agreement to Tsai's values \cite{TSAI74} (small circles).
\item[The inelastic contribution.]("meson production form factor").
We calculated the cross-sections of each of the 5~bins listed 
in tab.\ref{tab:num1} and found the contributions from $W_1$ negligible
compared to $W_2$. The sum is shown in fig.\ref{fig:res1}.
The $\Delta$-resonance ($1^{\mbox{st}}$~bin) is the dominant contribution with
some 50\% at small and more than 80\% at large angles.  This is
expected, since the strength $\tilde{W_2}$ is largest and $M_f$ is lowest.
The contributions of the other bins decrease with increasing $M_f$.
\item[The deep inelastic contribution.]
For small angles, $Q^2$ becomes less than $Q^2_0=1\,\mbox{GeV}^2$
and our simple para\-metrizations of $W_{1,2}$ break down.
This is approximately the case for $\theta < \theta_{min}=Q_0/E$ 
(independent of k). Therefore we start our calculation at $\theta_{min}$
as it is shown in fig.\ref{fig:res1} (solid line).
For these small angles, the other contributions are already dominant.
Furthermore, the condition $M_f > 2.6\,\mbox{GeV}$ has to be fulfilled,
when scaling is applicable. This means that the lower limit $x_0$ 
in eq.(\ref{equ:cross19}) is shifted to a somewhat larger value
\begin{equation}
  \label{equ:res4}
   x_{o}^{\prime}= 
   \frac{1/2(M_{f,min}^2 - M_N^2) + M_N E}{M_N k - 2 E k \sin^2 (\theta /2)}
\end{equation}
In general, this is a small (less than 10\%) effect, since the muon energy
region where $E$ is not much larger than the nucleon mass $M_N$ is already
excluded by applying $\theta_{min}$.
\end{description}
As it can be seen in fig.\ref{fig:res1} the deep inelastic contribution
is very important for large muon scattering angles.
%
%
\section{Conclusions}
\label{conclu}
In conclusion
we have presented a new practical approach to the Bethe Heitler process 
for large scattering angles at high energies. 
Of course, in an exact evaluation of the Bethe Heitler expression, as it
is given in \cite{TSAI74} (equ.(2.3)) the muon pole contribution is 
included. However this expression is to cumbersome and time consuming
for practical purposes. We have shown how to include the 
``deep inelastic contribution'' and
 show that it is important for 
relevant numerical examples. This extends the results of Tsai \cite{TSAI74}.
Among other things, such contributions are important for muon shielding
problems at future linear colliders.
%
%

%
%

\begin{thebibliography}{}
\bibitem{CDR97} R.Brinkmann et al.:
        \textsl{Conceptual Design of a 500~GeV $e^+ e^-$ Linear Collider
                with Integrated X-ray Laser Facility.} 
        DESY 1997-048, ECFA 1997-182
\bibitem{BAUR98} G.Baur, A.Leuschner, K.Tesch:
        \textsl{Muon doses at earth surface above the Linear Collider:
                Improved calculations.} 
        Internal Report, DESY D3-91, September 1998
\bibitem{LALI86} L.D.Landau, E.M.Lifschitz:
        \textsl{Lehrbuch der Theoretischen Physik.} 
         Vol.IV, Akademie-Verlag, Berlin, 1986
\bibitem{TSAI74} Y.S.Tsai:
        Rev.Mod.Phys. 46(1974)815
\bibitem{KIM73} K.J.Kim, Y.S.Tsai:
        Phys.Rev. D8(1973)3109
\bibitem{CHEN75} M.Chen, P.Zerwas:
        Phys.Rev. D12(1975)187
\bibitem{CHAN93} G.Chanfray et al.:
        Nucl. Phys. A556(1993)439
\bibitem{BFK73} V.N.Baier, V.S.Fadin, V.H.Khoze:
        Nucl. Phys. B65(1973)381
\bibitem{BAR68} W.Bartel, et al.:
        Phys.Lett. 28B(1968)148
\bibitem{PAR98} C.Caso et al.:
        \textsl{Review of Particle Properties.} 
        Eur.Phys.J. C3(1998)1
\bibitem{KEN91} H.W.Kendall:
        Rev.Mod.Phys. 63(1991)597        
\bibitem{OHN94} J.Ohnemus, T.F.Walsch, P.M.Zerwas:
        Phys.Lett. B328(1994)369
\bibitem{FRI91} J.I.Friedman:
        Rev.Mod.Phys.63(1991)615
\bibitem{AUB87} J.J.Aubert et al.:
        Nucl.Phys.B293(1987)740
\bibitem{FEY72} R.Feynman, 
        \textit{Photon-Hadron Interactions, Frontiers in Physics Series} 
        (edited by W.A.Benjamin,Inc1972) page 130
%
\end{thebibliography}
\end{document}